\documentclass[prd,nofootinbib,letterpaper,twocolumn,floatfix]{revtex4}
\usepackage{amsbsy}
\usepackage{latexsym}
\usepackage{graphicx}
\usepackage{color}
\usepackage{color}
\usepackage{times}
\usepackage{graphicx}
\usepackage{fancyhdr}
\usepackage{float}
\usepackage{ulem}
\makeatletter
\makeatletter
\def\@fnsymbol#1{\ifcase#1\or * \or  $+$ \or  \$ \or \#  \or \dag \or \ddag \or
$\mathsection$ \or $ \mathparagraph$ \or $\|$  \or \textordfeminine \or \textbul
let   
\or ** \or $++$ \or  \$\$ \or \#\#  \or \dag\dag \or \ddag\ddag \or
$\mathsection\mathsection$ \or $ \mathparagraph\mathparagraph$ \or $\|\|$  \or 
\textordfeminine\textordfeminine \or \textbullet \textbullet \or *** \or $+++$ 
\or  \$\$\$ \or \#\#  \or \dag\dag \or \ddag\ddag \or
$\mathsection \mathsection\mathsection$ \or $ \mathparagraph 
\mathparagraph\mathparagraph$ \or $\|\|\|$  \or 
\textordfeminine\textordfeminine\textordfeminine \or 
\textbullet\textbullet\textbullet \or \else \@ctrerr\fi}
\makeatother

\newcommand\xm{\ensuremath{\hat{x}}}
\newcommand\Lambdam{\ensuremath{\hat{\Lambda}}}
\newcommand\epsilonm{\ensuremath{\hat{\epsilon}}}
\newcommand\nm{\ensuremath{\hat{n}}}

\newcommand\xt{\ensuremath{x^\ast}}

\newcommand{\mubpb}{{\boldsymbol\mu}}
\newcommand{\thet}{{\boldsymbol\theta}}

\def\thercsid{\relax}
\def\rcsid#1{\def\next##1#1{\def\thercsid{##1}}\next}
\rcsid$Id: upper-limit.tex,v 1.93 2007/10/02 08:43:21 sfairhur Exp $

\renewcommand{\today}{\number\day\space\ifcase\month\or
  January\or February\or March\or April\or May\or June\or
  July\or August\or September\or October\or November\or December\fi
  \space\number\year}

\begin{document}

\title{The Loudest Event Statistic: General Formulation, Properties and
Applications}

\author{Rahul Biswas}
\affiliation{University of Wisconsin-Milwaukee, Milwaukee, WI  53201, USA}
\author{Patrick R. Brady}
\affiliation{University of Wisconsin-Milwaukee, Milwaukee, WI  53201, USA}
\author{Jolien D. E. Creighton}
\affiliation{University of Wisconsin-Milwaukee, Milwaukee, WI  53201, USA}
\author{Stephen Fairhurst}
\affiliation{University of Wisconsin-Milwaukee, Milwaukee, WI  53201, USA}
\affiliation{School of Physics and Astronomy, Cardiff University,
Cardiff, CF2 3YB, United Kingdom}
\affiliation{LIGO - California Institute of Technology,
Pasadena, CA  91125, USA}

\begin{abstract} The use of the loudest observed event to generate
statistical statements about rate and strength has become standard in
searches for gravitational waves from compact binaries and pulsars.
The Bayesian formulation of the method is generalized in this paper to
allow for uncertainties both in the background estimate and in the
properties of the population being constrained. The method is also
extended to allow rate interval construction. 
Finally, it is shown how to combine the
results from multiple experiments and a comparison is drawn between
the upper limit obtained in a single search and the upper limit
obtained by combining the results of two experiments each of half the
original duration. To illustrate this, we look at an example case,
motivated by the search for gravitational waves from binary inspiral.
\end{abstract}

\maketitle
\section{Introduction}
\label{sec:intro}

In daily life, we often estimate the birth rate, the rate of automobile
fatalities, or the rate of hurricanes in the Gulf. In these cases, it
is reasonably easy to determine when one event has occurred and so the
best estimate is usually taken to be the number of events divided by
the observation time. As physicists and astronomers, {we know this 
is a good estimator of the rate of an underlying Poisson process.}
In these cases, the ability to identify events with high confidence is 
central to the correctness of the rate estimate.

We can carry this method over to more complicated observational
situations by allowing for false positives in our identification of
events. Experiments are usually designed so that the rate of real
(foreground) events is higher than the rate of false positive
(background) events. Hence a good estimate of the rate is obtained by
counting the number of events per unit time, and making a small correction
to allow for the false positives. This is the typical experimental
method of estimating the rate.

In both physics and astronomy, it is common to search for very rare
events in large data sets and we rely heavily on statistical methods
to interpret these searches. In this paper, we discuss the
problem of estimating the rate of these rare events.  When real events
are very rare or very weak, it is important to revisit the reasoning
that underlies the standard approaches to estimating rates (and
indeed other parameters). Here, we explore the effects of
incorporating information about quality of observed events into the
estimate of event rate. One measure of quality might be the signal to
noise of the events; the louder an event, the more likely it is to be
signal. Of course, more complicated measures are also possible. We
simply require a rank ordering such that a larger quality implies the 
event is less likely to be background.

A popular method of incorporating quality information is to fix a
threshold, prior to looking at the data. The threshold is often chosen
to give an acceptable rate of background events in some qualitative
sense. Then, the upper limit is determined by counting the number of
events per unit time above the chosen threshold and making a
correction which allows for the background. Central to this method is
the prescription by which the final list of events are identified.

There are many alternative criteria that might be used to determine the
sample of events in an experiment. We consider using the loudest event
to estimate the rate. This method was first introduced in
gravitational-wave searches during the analysis of data from a prototype
instrument~\cite{Allen:1999yt};  the method was used to determine an
upper limit on the rate of binary neutron star mergers in the Galaxy.
Since then, the method has been used in a number of searches for
gravitational waves~\cite{LIGOS1iul, LIGOS2macho, LIGOS2iul, LIGOS2bbh,
Abbott:2003yq, Siemens:2006vk}.  More details of this method of
determining an upper limit are available in \cite{loudestGWDAW03}.
Related methods have been discussed in the context of particle physics
experiments by Cousins \cite{Cousins:1993gs} and
Yellin~\cite{Yellin:2002xd}.

In Sec.~\ref{sec:loudest}, we present a general formulation of the
loudest event statistic~\cite{Allen:1999yt,loudestGWDAW03}. We adopt
the Bayesian approach which gives a posterior distribution over
physical parameters based on the loudest event observed in an
experiment. To provide a concrete example, in Sec.~\ref{sec:rate} we
specialize to the case of a single unknown rate amplitude multiplying
a known distribution of events. Confidence intervals based on the
loudest event posterior are discussed in
Sec.~\ref{sec:intervals}.   The approach we take
is sometimes called a highest posterior density
interval~\cite{Yao:2006}.  It provides a unified approach giving an
upper limit for a loudest event that is due to noise with high
probability and a confidence interval (bounded away from zero) when
the loudest event is foreground with high probability. In real
experiments there are many systematic uncertainties; we discuss
marginalization over uncertainties in Sec.~\ref{sec:margin}. Finally, we
explain how to combine the results from multiple experiments using the
loudest event method and show that the resulting upper limit is
independent of the order of the experiments. This discussion also
leads naturally to an investigation of the effect on an upper limit if
a single search is split into two parts. In Sec.~\ref{sec:discussion}, we make
some general comments on the results obtained in this paper. In
Appendix~\ref{sec:cousins}, we consider the application of a
Feldman-Cousins unified approach to obtaining a frequentist upper
limit using the loudest observed events. A comparison between the
loudest event method and an event counting method of obtaining an
upper limit is given in Appendix~\ref{sec:comp} for a toy problem.

Notation: We use the following notation throughout this paper.
The loudest event statistic variable is denoted $x$.
The experimentally measured values of a quantity are denoted by
a circumflex accent; e.g., the experimentally measured value of
the loudest event statistic is given by $\xm$.  Probability distributions
and other quantities related to an experimental background appear
with a subscripted 0; e.g., $P_0(x)$ is the background probability
distribution for the loudest event statistic.  The symbol $B$ is used
to describe the collective information about the experimental background
in conditional probability distributions that are contingent on this
information; thus $P(x|B)$ is the distribution of the loudest event statistic
in an experiment that includes a background.

\section{General Formulation of Loudest Event Statistic}
\label{sec:loudest}

Consider a search of experimental data for a rare Poisson process. The output
of this search is a set of candidate events which have survived all cuts
applied during the analysis. At first, suppose that all the events are
\emph{foreground} events.  Assume that these events can be ranked
according to a single parameter $x$, such as a signal-to-noise ratio, in
such a way that the probability that the search will detect an event
increases with increasing $x$.  For simplicity we will call this parameter
the \emph{loudness} parameter and we will say that the event with the largest
value of $x$ is the \emph{loudest event}.  If the mean number of events
expected during the course of the experiment with the ranking statistic value
above $x$ is given by $\nu(x)$, then the probability of observing no events
above a given value of $x$ is
\begin{equation}\label{eq:fg_rate}
  P(x) = e^{-\nu(x)} . 
\end{equation}
However, if the experiment can produce \emph{background} events, then the
probability that there are no events, either foreground or background, 
louder than $x$ is
\begin{equation}\label{eq:rate_w_bg}
  P(x| B) = P_0(x) e^{-\nu(x)}
\end{equation}
where we have used $B$ to indicate that the probability depends on the
background and the factor $P_0(x)$ is the probability of obtaining zero
background events louder than $x$.

The mean number of events expected during the course of the experiment,
$\nu(x)$, depends on the duration of the experiment, the rate of events,
and the ability of the experiment to detect events that occur.  The
sensitivity of the search is encoded in the efficiency which is the probability
that an event will have a loudness value greater than or equal to $x$.
The efficiency depends on $x$ as well as a set of parameters, collectively
denoted by $\thet$, that determine the detectability of a source.
For example, $\thet$ may include such things as the sky position,
orientation, distance, etc., of an astrophysical source.
We write the efficiency as $\varepsilon(x,t,\thet)$.  (The sensitivity of the
experiment may change with time; hence the explicit dependence on $t$ in the
efficiency.)  The rate of events depends on the parameters $\thet$ that
describe the source population as well as on physical parameters, collectively
denoted by $\mubpb$, that we are interested in measuring
or constraining by means of the experiment.  We write the rate of events
as $R(\thet,\mubpb)$.  With this factorization, the mean number of events
expected can be expressed as
\begin{equation}\label{eq:rate_integral}
  \nu(x,\mubpb) = \int_0^T dt \int d\thet\, \varepsilon(x,t,\thet) 
  R(\thet,\mubpb) \, , 
\end{equation}
where $T$ is the total observation time.

We can substitute our expression for the rate (\ref{eq:rate_integral})
into Eq.~(\ref{eq:rate_w_bg}) to obtain the probability that
there are zero events in the data with a loudness statistic value greater than
$x$ as
\begin{widetext}
\begin{equation}\label{eq:fg_cum_dist}
  P(x|\mubpb, B) = P_0(x) \exp\left\{
  -\int_0^T dt \int d\thet\, \varepsilon(x,t,\thet) R(\thet,\mubpb) 
  \right\}.
\end{equation}
Furthermore, the probability of the loudest event occurring between
$x$ and $x+dx$ is given by $p(x|\mubpb,B)\,dx$ where
\begin{eqnarray}\label{eq:fg_bg_dist}
  p(x|\mubpb,B) &=& \frac{d}{dx} P(x|\mubpb,B) \nonumber \\
  &=& p_0(x)\left[ 1 - \left( \frac{P_0(x)}{p_0(x)} \right) 
  \int_0^T dt \int d\thet\, \frac{d\varepsilon(x,t,\thet)}{dx} 
  R(\thet,\mubpb) \right] e^{-\nu(x,\mubpb)}
  \label{eq:fg_bg_dist2}
\end{eqnarray}
\end{widetext}
and $p_0(x)=dP(x)/dx$.
Notice that the probability distribution contains two factors: an exponential
decay that is determined by the (foreground) rate of events and a shape factor
comprising two terms.  

After performing an experiment, we are interested in obtaining a distribution
for the model parameters that govern the rate.  To do this, we calculate
a Bayesian posterior distribution for these parameters, $\mubpb$, given the
observations.  This distribution is denoted $p(\mubpb|\xm,B)$,
where $\xm$ is the value of the observed loudest event, and
it is derived using Bayes' law:
\begin{equation}\label{eq:posterior}
  p(\mubpb|\xm,B) =
  \frac{p(\mubpb)\,p(\xm|\mubpb,B)}{\int d\mubpb\,p(\mubpb)\,p(\xm|\mubpb,B)}
\end{equation}
where $p(\mubpb)$ is the prior probability distribution on the model
parameters.  In many circumstances, the parameters $\mubpb$ may be
further divided into a set of particular interest $\mubpb_{\mathrm{I}}$
and others of less interest $\mubpb_{\mathrm{II}}$. By integrating
Eq.~(\ref{eq:posterior}) over the unwanted parameters
$\mubpb_{\mathrm{II}}$, one obtains the posterior distribution
\begin{equation}\label{eq:marginalized}
  p(\mubpb_{\mathrm{I}}|\xm,B) =
  \frac{\int d\mubpb_{\mathrm{II}}\, p(\mubpb_{\mathrm{I}}, 
  \mubpb_{\mathrm{II}})\,p(\xm|\mubpb_{\mathrm{I}},\mubpb_{\mathrm{II}},B)}
  {\int d\mubpb\,p(\mubpb)\,p(\xm|\mubpb,B)}.
\end{equation}
In Sec.~\ref{sec:margin} we consider this procedure of marginalization
over unwanted, or nuisance, parameters in more detail. 

To bound the parameters of interest at a given
confidence level $\alpha$, one integrates Eq.~(\ref{eq:marginalized})
over some region $\Omega(\mubpb_{\mathrm{I}})$ such that
\begin{equation}\label{eq:rate_limit}
  \alpha = \int_{\Omega(\mubpb_{\mathrm{I}})} d\mubpb \,
  p(\mubpb_{\mathrm{I}}|\xm,B).
\end{equation}
In general, the difficult part is selecting the region
$\Omega(\mubpb_{\mathrm{I}})$, especially in more than one dimension.
There are several ways to do this: for example, one could marginalize
over all but one of the parameters thus reducing the problem to a
one-dimensional integral; or select the smallest volume
$\Omega(\mubpb_{\mathrm{I}})$ that gives the required probability.
This is sometimes called a highest posterior density
interval~\cite{Yao:2006}. In Sec.~\ref{sec:intervals}, we investigate
the properties of this type of rate interval based on the loudest event
method.

\section{Upper limit on unknown rate amplitude}
\label{sec:rate}

We have obtained the general expression for the posterior probability
distribution of the parameters $\mubpb$ governing an astrophysical model
based on an observed loudest event. In practice, the details of
obtaining either a rate upper limit or a confidence interval on the
model parameters will depend upon the details of the astrophysical model
and its dependence upon the variables $\mubpb$.  In this section, we
simplify to the situation where the rate is dependent upon a single
parameter $\mu$, an overall unknown Poisson mean number of events, so that
\begin{equation}\label{eq:1d_mu}
  R(\thet,\mu) = \frac{\mu f(\thet)}{T}
\end{equation}
where $T$ is the observation time and $f(\thet)$ is the distribution of
events as a function of $\thet$.

We can use this form of the rate to simplify the general expression for
the posterior.  To begin, we introduce the quantity
\begin{equation}
  \epsilon(x) = \frac{1}{T}\int_0^T dt\int d\thet \, \varepsilon(x,t,\thet)
  f(\thet)
  \label{eq:efficiency}
\end{equation}
which can be regarded as an averaged detection efficiency: the
probability that a foreground event will have a loudness parameter
greater than $x$.  Then, the mean number of events with ranking
statistic above $x$ is $\nu(x) = \mu\epsilon(x)$, and (at least in
principle) $\epsilon(x)$ is known. The posterior distribution is
determined by substituting Eqs.~(\ref{eq:1d_mu}) into
(\ref{eq:rate_integral}) and using Eqs.~(\ref{eq:fg_bg_dist2}),
(\ref{eq:posterior}), and (\ref{eq:efficiency}) to obtain
\begin{equation}\label{eq:1d_post}
  p(\mu|\epsilonm,\Lambdam) =
  \frac{p(\mu) \, p_0(\xm) \, (1 + \mu\epsilonm\Lambdam)e^{-\mu\epsilonm}}%
  {\int d\mu \, p(\mu) \, p_0(\xm) \, 
  (1 + \mu\epsilonm\Lambdam)e^{-\mu\epsilonm}}
\end{equation}
where the function $\Lambda(x)$ is given by
\begin{equation}\label{eq:Lambda}
  \Lambda(x) = \left(\frac{-1}{\epsilon(x)}\frac{d\epsilon(x)}{dx}\right)
  \left(\frac{p_0(x)}{P_0(x)}\right)^{-1}
\end{equation}
and a hat over a function indicates evaluation at $\hat{x}$.
The quantity $\Lambda(x)$ is a measure of the relative probability of
detecting a single event with loudness parameter $x$ versus such an
event occurring due to the experimental background; in particular
$\Lambdam\to0$ in the limit that the
loudest event is definitely from the background and $\Lambdam\to\infty$ in
the limit that the loudest event is definitely from the foreground.  Note that
if the possibility that the loudest event could be from a background is
ignored, the posterior distribution, $p(\mu|\epsilonm) \propto \mu \, 
\epsilonm \, e^{-\mu\epsilonm}$ is peaked away from zero
and vanishes as $\mu\to0$; that is, the posterior distribution will be
inconsistent with zero foreground events.  

For the rest of the paper we take a uniform prior except in Section
\ref{sec:multiple} in which we consider using the posterior from a first
experiment as a prior for a second experiment.  It should be noted that
only power law priors on the rate, including the uniform prior, do not
introduce a timescale into the problem.  Power laws with powers greater
than -1 are needed to avoid a required low-rate cutoff, which again
would introduce a natural timescale.

Let us evaluate the upper limit making use of a uniform prior,
\begin{equation}\label{eq:uniform}
  p(\mu) = \mbox{const}.
\end{equation}
While this distribution is not normalizable, we can introduce a
cutoff at large $\mu$ (well above the expected number of events during the given
experiment) in order to render it normalizable.  Physically, this is a
reasonable choice of prior if there is no information available about
the expected value of $\mu$.  Furthermore, the posterior distribution
is insensitive to the value of the cutoff provided it is sufficiently large.
For the uniform prior, the posterior distribution in
Eq.~(\ref{eq:1d_post}) evaluates to
\begin{equation}\label{eq:post_uniform}
  p(\mu | \epsilonm, \Lambdam ) =  \frac{\epsilonm}{1+\Lambdam}
  (1 + \mu \epsilonm \Lambdam) e^{-\mu \epsilonm}.
\end{equation}

It is straightforward to show that the distribution in Eq.~(\ref{eq:1d_post})
will be peaked away from zero if and only if $\Lambdam > 1$;
the mode of the distribution is
\begin{equation}\label{eq:mode}
  \mu_{\mathrm{peak}} = \left\{
  \begin{array}{ll}
  0 & \quad \Lambdam \le 1 \\
  (\Lambdam - 1)/\Lambdam\epsilonm & \quad \Lambdam > 1.
  \end{array}
  \right.
\end{equation}
If $\Lambdam > 1$ then one might take this as an indication of a
non-zero rate. The extent to which this is true is explored in
Sec.~\ref{sec:intervals}.

\begin{figure}[t]
\includegraphics[width=1.0\linewidth]{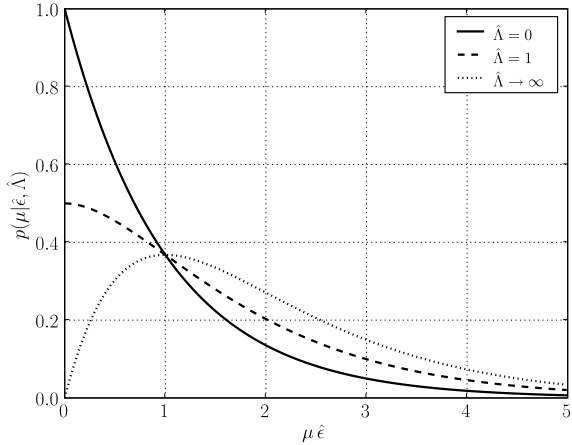}
\caption{\label{fig:rate_plot} The posterior probability density
function $p(\mu|\epsilonm,\Lambdam)$ on the Poisson mean $\mu$, assuming
a uniform prior, and a fixed value $\epsilonm$ of the efficiency
evaluated at the loudest event $\xm$.  The three curves correspond to
three different values of $\Lambdam$: a) $\Lambdam = 0$ (solid line),
the loudest event is definitely background and the distribution is
exponential; b) $\Lambdam = 1$ (dashed line), the transitional case
where the the distribution peaks at zero but the derivative vanishes
there; c) $\Lambdam\to\infty$ (dotted line), the loudest event is
definitely from the foreground, the distribution is peaked away from
zero.}
\end{figure}

We integrate Eq.~(\ref{eq:post_uniform}) to obtain an upper limit
at confidence level $\alpha$ by solving
\begin{eqnarray}\label{eq:bg_uniform_limit}
  \alpha &=& \int_0^{\mu} d\mu' p(\mu' | \hat{\epsilon},\hat{\Lambda})
  \nonumber \\
  &=& 1 - \left[ 1 + \frac{\mu \epsilonm \Lambdam}{1 + \Lambdam }
  \right] e^{-\mu\epsilon(\xm)}  
\end{eqnarray}
for $\mu$. It has been shown in \cite{loudestGWDAW03} that setting the
background to zero yields a conservative rate limit. In the Bayesian
analysis, however, this yields a posterior probability distribution
function which is peaked away from zero, and goes to zero at zero rate.
This is clearly seen in Fig.~\ref{fig:rate_plot} which shows the
posterior distribution for three values of $\Lambdam$ including $\Lambdam
\to \infty$.  This is not surprising as we have neglected the
background, in which case the existence of a loudest event implies a
non-zero rate.  Although this does not invalidate the upper limit,
 it does mean that the posterior would not serve as a
suitable prior for a future experiment, as it is inconsistent with a
zero rate.  Nevertheless, it is still possible to obtain the upper limit as
\begin{equation}
  \frac{\mu_{90\%}}{T} = \frac{3.890}{T\epsilon(\xm)} \;.
\end{equation}

Similarly, the no-foreground limit can be obtained by taking $\Lambdam = 0$.
In this case, the 90\% confidence limit tends to
\begin{equation}
  \frac{\mu_{90\%}}{T} = \frac{2.303}{T\epsilon(\xm)}.
\end{equation}
Finally, we can consider the transitional case $\Lambdam=1$:
\begin{equation}
  \frac{\mu_{90\%}}{T} = \frac{3.272}{T\epsilon(\xm)}.
\end{equation}
The posterior distribution for the Poisson mean $\mu$ for these three
possibilities is shown in Fig.~\ref{fig:rate_plot}.

\section{Confidence interval on unknown rate amplitude}
\label{sec:intervals}

In Sec.~\ref{sec:rate}, we derived the upper limit on the Poisson mean $\mu$
based on the loudest event.  However, in the case where the value of
$\Lambdam$ is large (likely to be foreground), one might prefer to
obtain a rate interval rather than an upper limit.  
For a uniform prior, the
mode $\mu_{\mathrm{peak}}$ of the posterior distribution for the Poisson mean,
given in Eq.~(\ref{eq:mode}), is non-zero whenever $\Lambdam > 1$.
Furthermore, in this case, $\mu_{\mathrm{peak}}$
asymptotes to $1/\epsilonm$ for large values of $\Lambdam$ as one might
expect.  How significant an indicator of a non-zero rate is having the
peak of rate distribution be non-zero?  In order to examine this
idea more precisely, we describe a method of constructing a rate
interval using the loudest event statistic which provides a unified
approach similar to Feldman and Cousins~\cite{Feldman:1997qc}.

At some confidence level $\alpha$, an interval is given by
$[\mu_1,\mu_2]$ such that 
\begin{equation}\label{eq:cl}
  \int_{\mu_1}^{\mu_2}p(\mu|\epsilonm,\Lambdam)\, d\mu = \alpha.
\end{equation}
A supplementary condition is required to select a unique interval: we
identify the interval which minimizes $|\mu_2-\mu_1|$ and contains the
mode of the distribution (or zero for $\Lambdam < 1$). This condition
clearly results in $\mu_1=0$ for small values of $\Lambdam$, i.e. when
the loudest event was likely to have arisen from the background, the
rate interval on the process we wish to constrain includes zero rate.

\begin{figure}[t]
\includegraphics[width=1.0\linewidth]{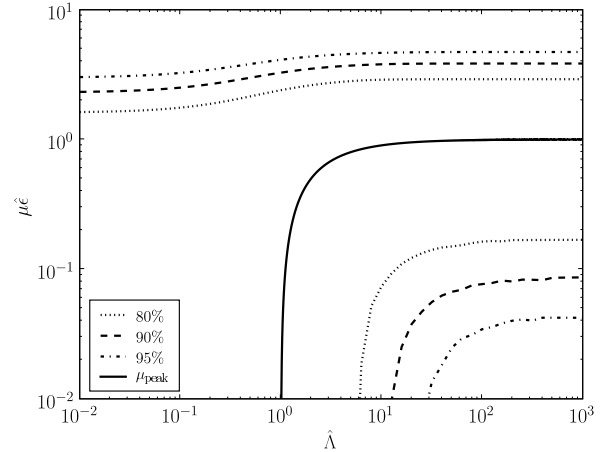}
\caption{\label{fig:small-interval} The graph shows the behavior of the
lower and upper boundaries of the interval, $\mu_{1}$ and $\mu_{2}$
respectively, as a function of $\Lambdam$.  They are
plotted for three different values of the confidence level $\alpha$ of
80\%, 90\% and 95\%.  The peak $\mu_{\mathrm{peak}}$ (solid line) 
approaches zero as
$\Lambdam$ approaches one.  As $\Lambdam \to 0$, $\mu_2$ agrees
with the no foreground upper limit treated above.}
\end{figure}

For the uniform prior, the dependence of $\mu_1$, $\mu_2$ and
$\mu_{\mathrm{peak}}$ on $\Lambdam$ are shown in
Fig.~\ref{fig:small-interval}. For $\Lambdam < 1$, $\mu_{\mathrm{peak}}
= 0$ and consequently $\mu_1=0$, as expected.  However, for a
significant range of $\Lambdam > 1$, even though the rate distribution
is peaked away from zero, $\mu_{1} = 0$ indicating that (at the given
confidence) the rate interval still includes zero.  

We can determine the precise value of $\Lambdam$ at which $\mu_1$ becomes
non-zero. For fixed $\Lambdam$ and $\epsilonm$, Eq.~(\ref{eq:cl}) gives
$\mu_2$ implicitly as a function of $\mu_1$. The minimal interval
condition is then just
\begin{equation}
\label{eq:derivative_cl}
\frac{d[\mu_2(\mu_1)-\mu_1]}{d\mu_1} = 0 .
\end{equation}
Substituting $\mu_1=0$ into Eqs.~(\ref{eq:cl}) and (\ref{eq:derivative_cl}), we obtain
two equations which depend on $\mu_2$ and $\Lambdam$. As an example,
consider a 90\% confidence interval.  In this case, $\mu_{1}$ becomes
non-zero, and the interval is bounded away from the origin,  at value of
$\Lambdam \simeq 11.56$.  This corresponds to $\mu_2 \simeq 3.807 /
\epsilon(\xm)$.  This result is in good  agreement with the values obtained
numerically in Fig.~\ref{fig:small-interval}.  

It is interesting to note that the 90\% confidence interval still
includes zero for a wide range of $\Lambdam$ that give posterior
distributions peaked away from zero. Figure~\ref{fig:margin_post} 
provides a concrete example of the posterior when $\Lambdam = 10$;
the 90\% confidence interval still includes zero.

\section{Marginalization over uncertainties}
\label{sec:margin}

The expected mean number of detected events, $\nu(x , \mubpb)$ in
Eq.~(\ref{eq:rate_integral}), is dependent upon the frequency of events
and their amplitude distribution as well as the sensitivity of the
search which is performed.  In many cases, neither of these quantities
will be precisely known.  For example, the efficiency of an experiment
is often measured via Monte-Carlo methods and therefore suffers from
uncertainties due to the finite number of trials.  If we expand our
understanding of the parameters $\mubpb$ to further parametrize the
uncertainties that can arise in the underlying models and in
measurements of efficiency, it is natural to marginalize over these
uncertainties before computing an upper limit or rate interval.  Just as
the marginalization over uninteresting physical parameters [given in
Eq.~(\ref{eq:marginalized})] requires a prior distribution to be
specified, the same is true of the uncertainties.  This prior
distribution would typically reflect the systematic and statistical
errors estimate for the experiment.

\subsection{Marginalization over uncertainties in $\epsilon$}
\label{sec:margin_epsilon}
As a particular example, consider the problem of the unknown rate
amplitude presented in Sec.~\ref{sec:rate} and assume there is some
uncertainty associated with the value of $\epsilonm=\epsilon(\xm)$.
Typically, one might choose the prior to be a normal distribution of the
variate $\epsilon$ peaked around the estimate value of $\epsilonm$. It
is, however, unphysical for the rate to be zero, so the distribution
would need to be truncated.  A more natural choice is a log-normal
distribution, for which the logarithm of $\epsilon$ would be normally
distributed, thereby guaranteeing that $\epsilon$ is positive. 

Here, we choose to make use of the $\gamma$-distribution, primarily
because it can be analytically integrated.  The $\gamma$-distribution is
similar in shape (for small standard deviation) to both the Gaussian and
log-normal distributions and in addition takes only non-negative
values.  The $\gamma$-distribution is given by 
\begin{equation}\label{eq:gamma}
  p(\epsilon ; k, \theta) = 
    \frac{\epsilon^{(k-1)} e^{-\epsilon/\theta}}{\theta^{k} \Gamma(k)}
\end{equation}
where $\Gamma(k)$ is the Gamma function.
The mean is $\bar{\epsilon}=k\theta$ while the standard deviation is
$\sigma_{\epsilon} = k^{1/2}\theta$.  Therefore, the fractional standard
deviation, $\sigma_{\epsilon}/\bar{\epsilon} = k^{-1/2}$ tends to zero
in the limit as $k \to \infty$, whereby we expect to recover the
unmarginalized results.  Note that the $\gamma$-distribution is a
distribution over the domain $\epsilon\in[0,\infty)$ while the efficiency
actually takes values only between 0 and 1.  If $k$ is large then the
$\gamma$-distribution is sharply peaked about its mean value and we can ignore
this issue.

The marginalized distribution is calculated by integrating over $\epsilon$,
\begin{equation}
  p(\mu | k, \epsilonm, \Lambdam ) = \left[ 
  \int_0^\infty d\epsilon \, p(\epsilon ; k, \theta) \, p(\mu | \epsilon, \Lambdam)
  \right]_{\theta=\epsilonm/k}.
\end{equation}
where we set the value of the parameter $\theta$ so that the mean of
the $\gamma$-distribution equals the observed efficiency $\epsilonm$,
and where $k$ is a measure of the fractional uncertainty in the value
of $\epsilonm$.  Making use of the distribution (\ref{eq:1d_post}) for the
Poisson mean parameter and the expression for the $\gamma$-distribution given
above, we obtain the marginalized distribution for integer values of $k$
\begin{widetext}
\begin{equation}\label{eq:margin_post}
  p(\mu | k, \epsilonm, \Lambdam ) = 
  \frac{ \epsilonm }{( 1 + \Lambdam)} \left[
  \frac{1}{\left( 1 + \mu \epsilonm/k \right)^{k + 1}} +
  \frac{\mu\epsilonm\Lambdam(1+1/k)}{(1+\mu\epsilonm/k)^{k+2}}
  \right].
\end{equation}
\end{widetext}
In the limit that $k \to \infty$, we recover the previous
distribution for $\mu$ as expected. 

\begin{figure}[t]
\includegraphics[width=1.0\linewidth]{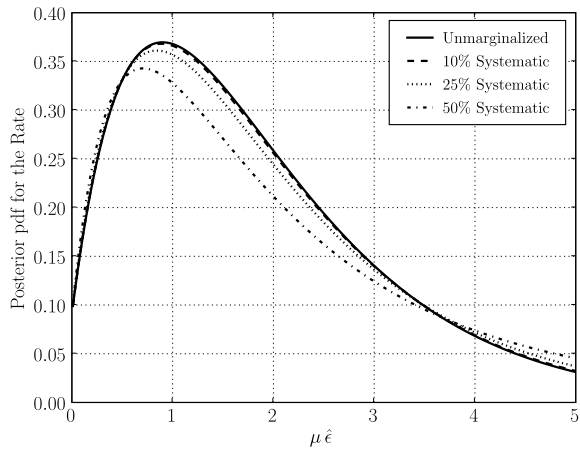}
\caption{\label{fig:margin_post} The posterior probability density
function on the Poisson mean parameter $\mu$ for different sizes of
systematic error, where $\epsilonm$ is the efficiency evaluated at the
loudest event $\xm$.  The curves were generated assuming a uniform prior
and using $\Lambdam = 10$.  The solid line corresponds to the
unmarginalized probability density function. The dot-dashed line gives
the distribution marginalized over a $10\%$ systematic uncertainty
(equivalently $k = 100$ for the $\gamma$-distribution).  With this level
of uncertainty, the marginalized distribution is barely changed from the
original.  The dotted and dashed lines show the posterior for $25\%$ ($k
= 16$) and $50\%$ ($k=4$) systematic errors.  As the systematic error
increases the distribution broadens and consequently the upper limit
increases.}
\end{figure}

\begin{figure}[t]
\includegraphics[width=1.0\linewidth]{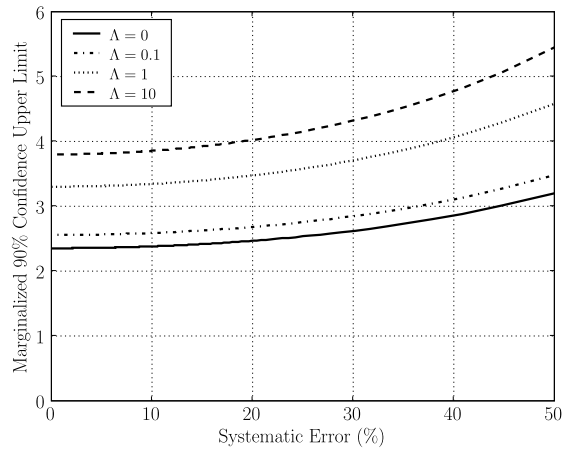}
\caption{\label{fig:rate_vs_k} The $90\%$ confidence upper limit versus
the size of the systematic error which is marginalized over (equivalent
to $k^{-1/2}$ in the $\gamma$-distribution discussed in the text).
The limit is plotted for four different values of $\Lambdam$: $0, 0.1, 1,
10$.  In all cases, the upper limit increases with larger systematic
error.}
\end{figure}

In order to examine the effect of marginalization, in
Fig.~\ref{fig:margin_post} we plot the unmarginalized posterior distribution
for $\Lambdam = 10$ along with three distributions obtained by
marginalizing over different size systematic errors or uncertainties.
These distributions are obtained from (\ref{eq:margin_post}) with values
of $k = 100$, $16$ and $4$ corresponding to errors of $10\%$, $25\%$ and $50\%$
respectively.  As the systematic error increases, the posterior distribution
for the Poisson mean parameter gets broader; the value of the probability
density function increases for large values of the Poisson mean parameter.
This causes an increase in the upper limit.  Without taking into account any
uncertainties, the $90\%$
confidence upper limit is $3.796/\epsilon(\xm)$.  For $10\%$ systematic
error, this increases only slightly to $3.850/\epsilon(\xm)$ while for $25\%$
and $50\%$ this increases further to $4.147/\epsilon(\xm)$ and
$5.434/\epsilon(\xm)$ respectively.  In Figure \ref{fig:rate_vs_k} we plot
the upper limit as a function of the systematic error for four different
values of $\Lambda$.  The results are qualitatively similar to what was
seen before --- marginalizing over uncertainties will increase the upper
limit and the larger the errors, the larger the effect.

\subsection{Marginalization over uncertainties in $\Lambda$}
\label{sec:margin_mu}
In many cases, there will also be uncertainties in the precise value of
$\Lambdam=\Lambda(\xm)$.  These can be marginalized over in the same way as described
above.  Since the $\Lambdam$ dependence of the distribution
(\ref{eq:1d_post}) is straightforward, this can be done
explicitly. For concreteness, let us take a uniform prior, in
which case the posterior distribution is given by Eq.~(\ref{eq:post_uniform}).
Then, given a probability distribution $p(\Lambda)$, the marginalized
distribution is
\begin{equation}\label{eq:mu_margin}
  p(\mu | \epsilonm) = \int d\Lambda \, p(\Lambda) \, 
  p(\mu | \epsilonm, \Lambda) 
\end{equation}
In this case, the above integral is straightforward.  Specifically,
let us define
\begin{equation}\label{eq:xi}
  \xi = \int d\Lambda \, p(\Lambda) \, \frac{\Lambda}{(1 + \Lambda)} .
\end{equation}
Then, the posterior distribution following marginalization over
$\Lambda$ is given by
\begin{equation}\label{eq:mu_post}
  p(\mu | \epsilonm ) = \epsilonm
  \left[(1-\xi) + \mu \epsilonm \xi\right]
  e^{-\mu \epsilonm}
\end{equation}
where $\xi$ contains all of the dependence of the posterior on the
marginalized background.

Suppose that $\Lambda$ is distributed with expectation value
$\Lambdam$ and variance $\sigma^{2}_{\Lambda}$.  Then, to leading
order, 
\begin{equation}\label{eq:xi_val}
  \xi \approx 
  \frac{\Lambdam}{1 + \Lambdam} - 
  \frac{\sigma_{\Lambda}^{2}}{(1 + \Lambdam)^{3}} .
\end{equation}
From this, we notice two things.  First, even if the fractional
uncertainties in $\Lambdam$ are of order unity, when $\Lambdam \gg 1$ or
$\Lambdam \ll 1$, the second term is small compared to the first and can
be ignored.  Second, marginalizing over $\Lambda$ only serves to
decrease the value of $\xi$ relative to the unmarginalized case.  This
is equivalent to reducing the likelihood that the loudest event is
foreground and consequently will reduce the upper limit.  Therefore, it
is possible to neglect the marginalization of $\Lambda$ as this is a
conservative thing to do.

\section{Combining results from multiple experiments}
\label{sec:multiple}

When performing a series of experiments, there is a very natural way to
combine the results in a Bayesian manner.  As discussed above, the
calculation of a Bayesian upper limit requires the specification of a
prior probability distribution for the rate $\mu$.
When a previous experiment has been performed, it is natural to use the
posterior from the first experiment as the prior for the second. It is
straightforward to show that the results are independent of the order of the
experiments.  (This does not depend upon the loudest event, rather it is a
general Bayesian result.)  Begin by recalling that
\begin{equation}
  p(\mu | \xm_1 ) = \frac{ p(\mu) \, p(\xm_1 | \mu) }
  { \int d\mu \, p(\mu) \, p(\xm_1 | \mu ) }.
\end{equation}
For the second search, simply use $p(\mu | \xm_1 )$ as the
prior to obtain the posterior distribution on $\mu$ given the
observations in both the first and second experiments:
\begin{equation}\label{eq:splitpost}
  p(\mu | \xm_1, \xm_2) = \frac{ 
    p(\mu) \, p(\xm_1 | \mu) \, p(\xm_2 | \mu) }
    { \int d\mu \, p(\mu) \, p(\xm_1 | \mu) \, p(\xm_2 | \mu) }.
\end{equation}
This is clearly symmetric in $\xm_1$ and $\xm_2$.  It is 
straightforward to see that marginalization over nuisance parameters
(see Sec.~\ref{sec:margin}) preserves this symmetry.

Let us consider this in more detail.  If the first
search was performed using a uniform prior, the posterior is given by
Eq.~(\ref{eq:1d_post}) with a loudest event value $\xm_1$ observed in the
first search.  Furthermore, in the event that the loudest event is
most likely background, one expects $\Lambdam_1 \ll 1$.  Then, we can 
conservatively rewrite the posterior as
\begin{equation}\label{eq:conservative}
  p_{\mathrm{conservative}}(\mu|\epsilonm_1,\Lambdam_1) =
  \epsilonm_1 \Lambdam_1 e^{-\mu\epsilonm_1(1-\Lambdam_1)} 
\end{equation}
where $\epsilonm_1=\epsilon_1 (\xm_1)$, $\Lambdam_1=\Lambda_1(\xm_1)$, 
and we have made use of the fact that 
\begin{equation}
  1 + \mu\epsilonm_1\Lambdam_1 \le e^{\mu\epsilonm_1\Lambdam_1}.
\end{equation}
It is straightforward to
show that the rate limit at a given confidence level $\alpha$ inferred using
this posterior is necessarily larger than that obtained using the original
distribution.  In this sense, the alternative distribution is conservative and
the distribution has been cast as an exponential.

Therefore, in the second search, it is natural to use an exponential
prior, 
\begin{equation}
  p(\mu) =\kappa e^{-\kappa\mu}.
\end{equation}
To obtain the posterior distribution obtained when the exponential
prior is used, it is beneficial to re-define $\Lambda(x)$ as
\begin{equation}\label{eq:exp_mu}
  \Lambda_\kappa(x) = \left(\frac{-1}{\epsilon_\kappa(x)}
  \frac{d\epsilon_\kappa(x)}{dx}\right)
  \left(\frac{p_0(x)}{P_0(x)}\right)^{-1}
\end{equation}
where
\begin{equation}
  \epsilon_\kappa(x) = \epsilon(x) + \kappa
\end{equation}
includes the exponential scale constant from the prior distribution.
Then, the posterior distribution is given by 
\begin{equation}\label{eq:bg_posterior}
  p(\mu|\epsilonm_\kappa,\Lambdam_\kappa) \propto 
  (1 + \mu\epsilonm_\kappa\Lambdam_\kappa) e^{-\mu\epsilonm_\kappa}
\end{equation}
with $\epsilonm_\kappa= \epsilon_\kappa(\xm)$ and $\Lambdam_\kappa=
\Lambda_\kappa(\xm)$.  As before, the posterior distribution is peaked
away from zero if $\Lambdam_\kappa > 1$.  In addition, the distribution
is identical to that obtained using a uniform prior, only now the search
efficiency is effectively $\epsilon(\xm) + \kappa$.

\subsection{Splitting a search}
Next, let us consider the effect of taking a single search and splitting
it into two halves, which can be combined to produce an upper limit in
the manner described above.  Naively, it appears that splitting the
search will give a lower rate limit, since we will be using a
quieter loudest event for half the search.  If this were the case,
then it would seem that splitting the search into ever shorter searches
would lower the  upper limit indefinitely.  As we shall see, the
result is not so clear cut, and it depends critically upon the
foreground and background distributions $\epsilon(x)$ and $P_0(x)$.

Consider an experiment performed for some given time $T$, and assume
that both the foreground and background rates are constant over time.
We would then like to compare the (expected) upper limit from the full
search to that obtained by splitting the data in two parts of length
$T_{1}$ and $T_{2}$ and calculating a combined upper limit from the two
searches. Let us assume, without loss of generality, that the loudest
event overall in the search occurs in the first half of the search with
a statistic value of $\xm_1$, and the loudest event in the second half
of the search has a statistic value $\xm_2$.  Then, we can calculate the
upper limit from the search (taking it as a single entity) and from the
split search.  

The posterior for the single search is given by
\begin{equation}
  p(\mu|\xm_1, B) =
  \frac{p(\mu)\left[1 + \mu \epsilon(\xm_1) \Lambda(\xm_1) \right] 
  \exp\{-\mu \epsilon(\xm_1) \}}%
  {\int d\mu\,p(\mu) \left[1 + \mu \epsilon(\xm_1) \Lambda(\xm_1)\right]\  
  \exp\{-\mu \epsilon(\xm_1) \}}
\end{equation}
while for the split search, the likelihood for each part is proportional to
\begin{equation}
  p(\xm_{i} | \mu, B) \propto
  \left[1 + \mu \eta_{i}\epsilonm_{i} \Lambdam_{i} \right] 
  e^{-\mu \eta_{i}\epsilonm_{i}} 
\end{equation}
where $i=1,2$ label the two parts of the search, $\epsilonm_i =
\epsilon_i(\xm_i)$, $\Lambdam_i = \Lambda_i(\xm_i)$ and $\eta_i = T_i/T$
is the fraction of the total observation time that is contained in the
each interval; $\eta_1+\eta_2=1$. Then the combined posterior distribution for the split search is
\begin{widetext}
\begin{equation}
p(\mu|\xm_1,\xm_2,B) = \frac{
p(\mu) \left[1 + \mu \eta_1 \epsilonm_1 \Lambdam_1 \right]
\left[1 + \mu \eta_2 \epsilonm_2 \Lambdam_2 \right]
\exp\{-\mu [\eta_1 \epsilonm_1 + \eta_2 \epsilonm_2 ] \}
}{
\int d\mu\,p(\mu) \left[1 + \mu \eta_1 \epsilonm_1 \Lambdam_1 \right]
\left[1 + \mu \eta_2 \epsilonm_2 \Lambdam_2 \right]
\exp\{-\mu [\eta_1 \epsilonm_1 + \eta_2 \epsilonm_2 ] \}
} .
\end{equation}
\end{widetext}

We can now compare the posterior distributions for the single and split
search. In general, the efficiencies $\epsilon_i(x)$ and the
likelihoods $\Lambda_i(x)$ could be different for the two parts.
Nevertheless, it
follows directly from Eq.~(\ref{eq:efficiency}) that
$\epsilon(x) = \eta_{1} \epsilon_{1}(x) + \eta_{2} \epsilon_{2}(x)$.
Therefore, in the split search, the exponential decay term is at least
as large as for the single search, with equality only if $x_2 = x_1$.
This tends to make the upper limit obtained in the split search
smaller than that of the single search.  In contrast, the polynomial
prefactor is always more significant for the split search (i.e. it
grows more steeply with $\mu$). This tends to make the upper limit
larger. So splitting the search will lead to a larger limit when
$\epsilonm_2 = \epsilonm_1$ and $\Lambdam_2 = \Lambdam_1$.  Meanwhile
if $\epsilonm_2 \gg \epsilonm_1$ and $\Lambdam_2 \ll \Lambdam_1$, the
split search will give a numerically smaller limit.

\begin{figure}[t]
\includegraphics[width=1.0\linewidth]{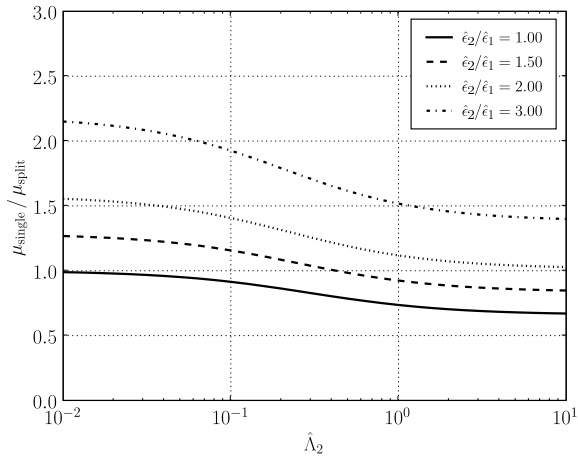}
\caption{\label{fig:single-split-comparison} 
The ratio of $\mu_{\mathrm{single}}$ to $\mu_{\mathrm{split}}$ as a
function of $\Lambdam_2$ for several values of
$\epsilonm_2/\epsilonm_1$.  The data presented uses a uniform prior on
$\mu$, $\eta_i=\frac12$, $\epsilon(x)=\epsilon_i(x)$ and $\Lambda_1(x)
= \Lambda_2(x)$.  The figure was generated for $\Lambda(\xm_1) = 0.5
\Lambda_1(\xm_1) = 1$.  In general, there is only a weak dependence on
this value; the curves steepen a little for smaller value of
$\Lambda(\xm_1)$, but look qualitatively similar. Note also that for
most sensible choices of amplitude statistic $x$, one expects
$\Lambdam_2 \leq \Lambdam_1$. The plot is extended to $\Lambdam_2 =10$
for completeness.}
\end{figure}

Assuming a uniform prior on $\mu$, we compare the $90\%$-confidence
upper limits on the Poisson mean obtained from a single search with
the limit obtained from a split search with $\eta_1 = \eta_2 = 1/2$.
The results are shown in Fig.~\ref{fig:single-split-comparison} for
several choices of $\epsilonm_2 / \epsilonm_1$ under the assumptions
that $\epsilon(x) = \epsilon_i(x)$ and $\Lambda_1(x) = \Lambda_2(x)$.
While not the most general case, these assumptions are reasonable in
the context of an experiment with the same apparatus and background
noise sources. In the limit
$\Lambdam_2 \rightarrow 0$, we find $\mu_{\mathrm{single}} >
\mu_{\mathrm{split}}$ as expected. As $\Lambdam_2$ increases, the
second event is less likely to be background and so the rate limit
from the split search can become bigger than that obtained in the
single search. While this makes intuitive sense, the result depends 
on the particular observed outcomes of the experiment.

When $\Lambdam_1 \ll 1$ and $\Lambdam_2 \ll 1$,
the posterior distribution for the single search can be approximated
conservatively as
\begin{equation}
  p(\mu|\xm_1, B) \simeq 
  \epsilon(\xm_1) [ 1 - \Lambda(\xm_1) ]
  e^{-\mu \epsilon(\xm_1) [ 1 - \Lambda(\xm_1) ]} 
\end{equation}
while the posterior for the split search becomes
\begin{equation}
  p(\mu|\xm_1,\xm_2, B) \simeq c(\xm_1,\xm_2)
  e^{-\mu c(\xm_1,\xm_2)} 
\end{equation}
where 
\begin{equation}
c(x_1,x_2) = \epsilonm_1 \eta_1 ( 1 - \Lambdam_1 ) 
+ \epsilonm_2 \eta_2 ( 1 - \Lambdam_2 ).
\end{equation}
If we further assume that both the foreground and background are
Poisson distributed over the entire search, then 
%
$\epsilon_{i}(x) = \epsilon(x)$ %
and 
$\eta_{i} \Lambda_{i}(x) = \Lambda(x)$.
%
Within the context of these assumptions, it is then easy to write
down the upper limit for each distribution.  In particular, 
\begin{equation}
  \mu_{\mathrm{single}} = 
  \frac{ - \ln(1-\alpha) }{ \epsilon(\xm_1) [ 1 - \Lambda(\xm_1) ] }
\end{equation}
for the single search; for the split search
\begin{equation}
  \mu_{\mathrm{split}} = \frac{ - \ln(1-\alpha) }{ c(\xm_1,\xm_2) }.
\end{equation}
Hence, the single search will give a smaller upper limit if
\begin{equation}\label{eq:rate_diff}
\Lambda_2(\xm_2) > 
\left[1 - \frac{\epsilon(\xm_1)}{\epsilon(\xm_2)}\right] \, .
\end{equation}
Once again, the comparison between the single and split search is
sensitive to the precise nature of the foreground, background and
observed results.

\section{Discussion}
\label{sec:discussion}

The loudest event statistic is just one method of taking account of
the quality of an event in the interpretation of a search. In this
paper, we have presented further exploration of the method including
the discussion of marginalization over uncertainties in the input
model. The Bayesian approach allows simple accounting of these
uncertainties by integrating them out. 

We also showed how the method could be used to determine a
rate interval. Once again, this is not the most powerful method of
determining an interval (in the sense that using more than one event
would lead to a more strongly peaked distribution and, consequently, a
narrower interval).  Nevertheless, the approach shows that a rate
interval arises when the likelihood that the event is signal becomes
large enough.  

Finally, we presented a discussion of combining the results from
multiple searches to determine a single upper limit. It was shown that
the limit obtained by combining two searches of equal duration is, in
general, different to the limit obtained by performing a single search
of equivalent duration. What conclusion to draw from this is unclear
since the notion of better depends on the true value of the rate being
explored.

Even though physicists have a deep appreciation for probabilistic
phenomena in nature, it is often tempting to talk about better upper
limits by using one method or another. This is, of course, a flawed
approach. In fact, it is the experiment that one should choose not the
statistical method. Nevertheless, some experiments may be more
powerful than others. For example, it would be ill-conceived to use
the loudest event method to determine a rate interval in an experiment
which is likely (in the sense of prior probability) to generate more
than one loud event that could be considered to arise from the
phenomenon of interest. Indeed, these considerations lead back to an
experiment more like the standard threshold approach.

\section*{Acknowledgments}

We would like to acknowledge many useful discussions with members of the
LIGO Scientific Collaboration inspiral analysis group which were
critical in the formulation of the methods and results described in this
paper. This work has been supported in part by NSF grants
PHY-0200852 and PHY-0701817; PRB is grateful to the Research
Corporation for support by a Cottrell Scholar 
Award; SF was funded in part by the
Royal Society. LIGO was constructed by the California Institute of
Technology and Massachusetts Institute of Technology with funding from
the National Science Foundation and operates under cooperative agreement
PHY-0107417.  This paper has LIGO Document Number LIGO-P070076-00-Z.

\appendix
\section{Frequentist approach to the loudest event statistic}
\label{sec:cousins}

\begin{figure}[t]
\includegraphics[width=1.0\linewidth]{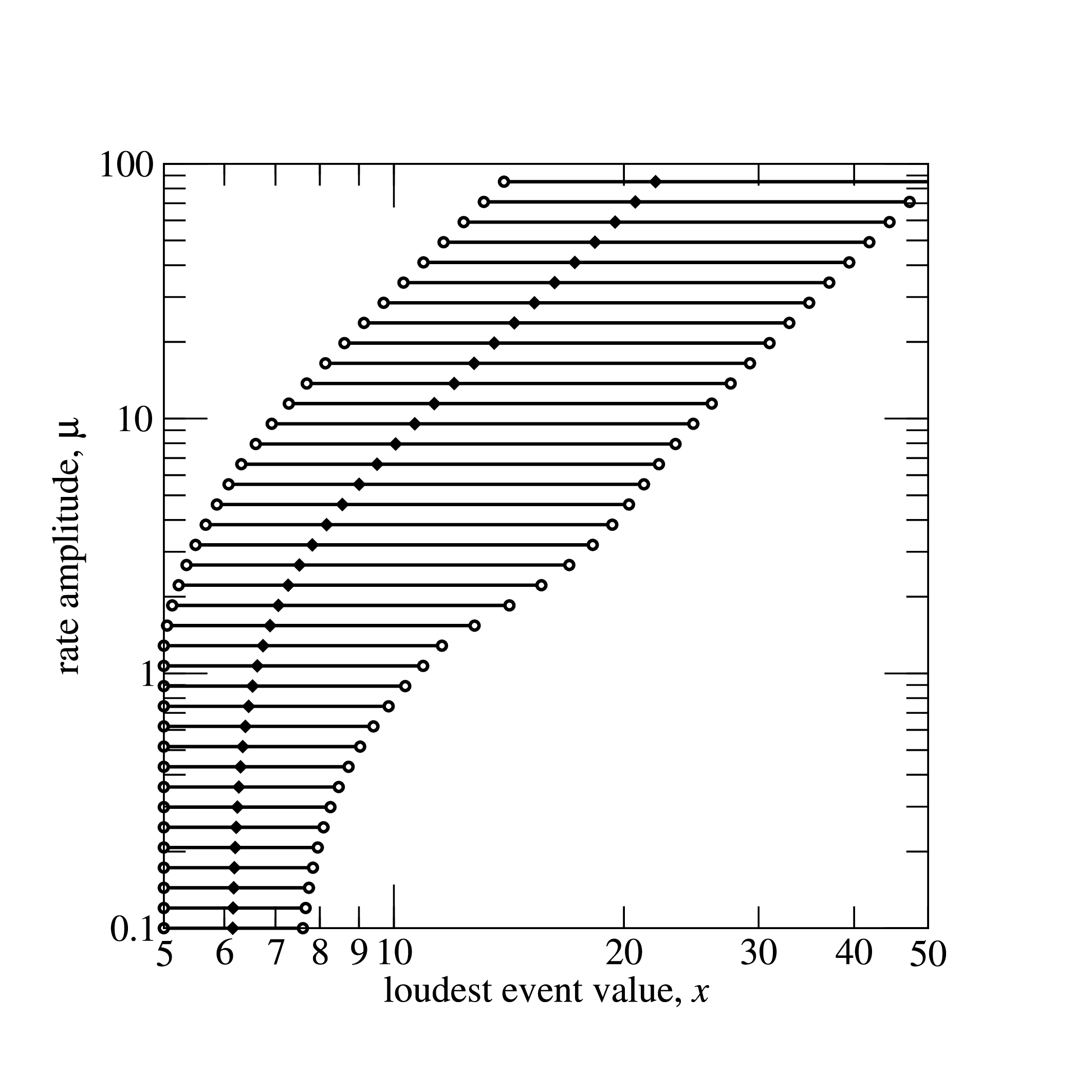}
\caption{\label{fig:belt}%
The $\alpha=90\%$ confidence belt constructed using the procedure of Feldman
and Cousins for the example described in the text.  For each value of $\mu$,
the probability of obtaining a loudest event value of $x$ in the interval
given by the solid line bounded by open circles is $90\%$.  The solid
diamond shows the point in this interval where $R(x)$ is the greatest.
For an observed loudest event value $\xm$, the interval on $\mu$ is the
intersection of the vertical line $x=\xm$ with this belt.  Note that for small
values of $\xm$ an upper limit on $\mu$ is obtained while for larger values of
$\xm$ the interval on $\mu$ is bounded away from zero.  The division between an
upper limit and an inteval that excludes zero occurs at a value of $x_{90\%}$
for which $P_0(x_{90\%})=90\%$.}
\end{figure}

This paper has explored the loudest event statistic from the Bayesian
point of view.  In this appendix we consider an alternative approach:
the construction of frequentist confidence intervals using the loudest
event statistic.  The construction of frequentist upper limits is
almost trivial (see~\cite{loudestGWDAW03}).  More interesting is the
application of the method of Feldman and Cousins \cite{Feldman:1997qc}
for a unified approach to constructing confidence intervals.  We restrict
attention here to the case in which only the unknown rate amplitude $\mu$
is to be bounded.

We briefly summarize the Neyman approach to constructing confidence belts: For
each fixed value of $\mu$, an interval $[x_1,x_2]$ is constructed such
that the probability of observing a loudest event $x$ in this interval
is equal to the desired confidence level $\alpha$:
\begin{equation}
  P(x_2|\mu) - P(x_1|\mu) = \alpha
\end{equation}
where
\begin{equation}
  P(x|\mu) = P_0(x)e^{-\mu\epsilon(x)}.
\end{equation}
The collection of such intervals then defines a confidence belt; for
any observed value of the loudest event $\xm$, the belt covers a range
of values of $\mu$, $[\mu_1,\mu_2]$, which is the desired interval on
the rate amplitude parameter.  In order to construct the confidence belt,
a supplementary condition is needed in order to uniquely define the
interval.  For example, to obtain a confidence belt that always yields upper
limits, choose $x_2=\infty$.  Then the belt is defined as the interval
$[x_1,\infty)$ where $x_1$ satisfies
\begin{equation}
  1 - P(x_1|\mu) = \alpha
\end{equation}
for each value of $\mu$.  Then, given an observed loudest event value $\xm$,
the rate amplitude interval is $[0,\mu_2]$ where
\begin{equation}
  1 - P(\xm|\mu_2) = \alpha
\end{equation}
or
\begin{equation}
  \mu_2 = - \frac{\ln(1-\alpha) - \ln P_0(\xm)}{\epsilon(\xm)}.
\end{equation}
As mentioned in \cite{loudestGWDAW03}, this procedure has the pathology
that if $P_0(\xm)<1-\alpha$ then the interval on $\mu$ is empty.

The unified approach of Feldman and Cousins provides a different
supplementary condition for defining the interval $[x_1,x_2]$ for
fixed $\mu$.  In the Feldman and Cousins approach, the interval is
constructed using a function $R(x)$ so that $R(x)\ge R_{\mathrm{min}}$
for $x\in[x_1,x_2]$ and $R(x)<R_{\mathrm{min}}$ for $x$ outside the
interval where $R_{\mathrm{min}}=\min\{R(x_1),R(x_2)\}$.  The function
$R(x)$ is chosen to be the likelihood ratio
\begin{eqnarray}
  R(x) &=& \frac{p(x|\mu)}{p(x|\mu_{\mathrm{peak}})} \\
  &=& \left\{ \begin{array}{ll}\displaystyle
  [1 + \mu\epsilon(x)\Lambda(x)]e^{-\mu\epsilon(x)} & \Lambda(x) \le 1 \\[\medskipamount]
  \displaystyle
  \frac{[1 + \mu\epsilon(x)\Lambda(x)]e^{-\mu\epsilon(x)}}{\Lambda(x)e^{1/\Lambda(x) - 1}} & \Lambda(x) > 1
  \end{array} \right. 
\end{eqnarray}
where
\begin{equation}
  p(x|\mu) = \frac{dP(x|\mu)}{dx} = p_0(x)[1+\mu\epsilon(x)\Lambda(x)]e^{-\mu\epsilon(x)}
\end{equation}
and $\mu_{\mathrm{peak}}$ is given by Eq.~(\ref{eq:mode}).

As an illustration, we compute the confidence belt according to the Feldman
and Cousins procedure for the following example:  We take
$\epsilon(x)=(x_{\mathrm{min}}/x)^3$ and $P_0(x)=1-\exp(x_{\mathrm{min}} - x)$
with $x_{\mathrm{min}}=5$; only values $x>x_{\mathrm{min}}$ are realizable.
The $\alpha=90\%$ confidence belt is shown in Fig.~\ref{fig:belt}.  Notice that
there is a well-defined inteval on $\mu$ for any observed loudest event value
$\xm$ possible, that is, the Feldman and Cousins procedure produces confidence
belts that are free of the pathology described above when upper limit
confidence belts are constructed.  Furthermore, the interval on $\mu$ is a
upper limit for small loudest event values $\xm<x_{90\%}$, but becomes an
interval which excludes zero for $\xm>x_{90\%}$ where $P_0(x_{90\%})=90\%$.

It is interesting to consider the behavior of the confidence intervals for
large values of $x$.  In this regime, $\Lambda(x)\gg1$, $P_0(x)\simeq1$ so
$R(x)\simeq\mu\epsilon(x)e^{-\mu\epsilon(x)+1}$ and $P(x|\mu)\simeq
e^{-\mu\epsilon(x)}$.  The confidence belt at fixed $\mu$ is given by the
interval $[x_1,x_2]$ that satisfies $R(x_1)=R(x_2)$ and
$P(x_2|\mu)-P(x_1|\mu)=\alpha$.  Therefore, $x_1$ and $x_2$ satisfy the coupled
equations
$\mu\epsilon(x_1)e^{-\mu\epsilon(x_1)}=\mu\epsilon(x_2)e^{-\mu\epsilon(x_2)}$
and $e^{-\mu\epsilon(x_2)}-e^{-\mu\epsilon(x_1)}=\alpha$.  For $\alpha=90\%$,
$\mu\epsilon(x_1)=3.932$ and $\mu\epsilon(x_2)=0.08381$.  Consequently, the
$90\%$ confidence rate amplitude interval $[\mu_1,\mu_2]$ for large $\xm$ will
be given by $\mu_1=0.08381/\epsilon(\xm)$ and $\mu_2=3.932/\epsilon(\xm)$.
Notice the ratio $\mu_2/\mu_1=46.91$ is fixed: the fractional uncertainty in
the value of $\mu$ does not improve as $\xm$ increases.  This demonstrates that
for experiments in which events in the low-background region are expected,
the loudest event statistic will not give strong constraints on the event
rate.  As emphasized in the introduction, the loudest event statistic is best
suited to problems in which the anticipated event rate is very low and
foreground and background events are expected to have comparable amplitudes.

\section{Comparison with Fixed Thresholds}
\label{sec:comp}

\begin{figure}[t]
\includegraphics[width=1.0\linewidth]{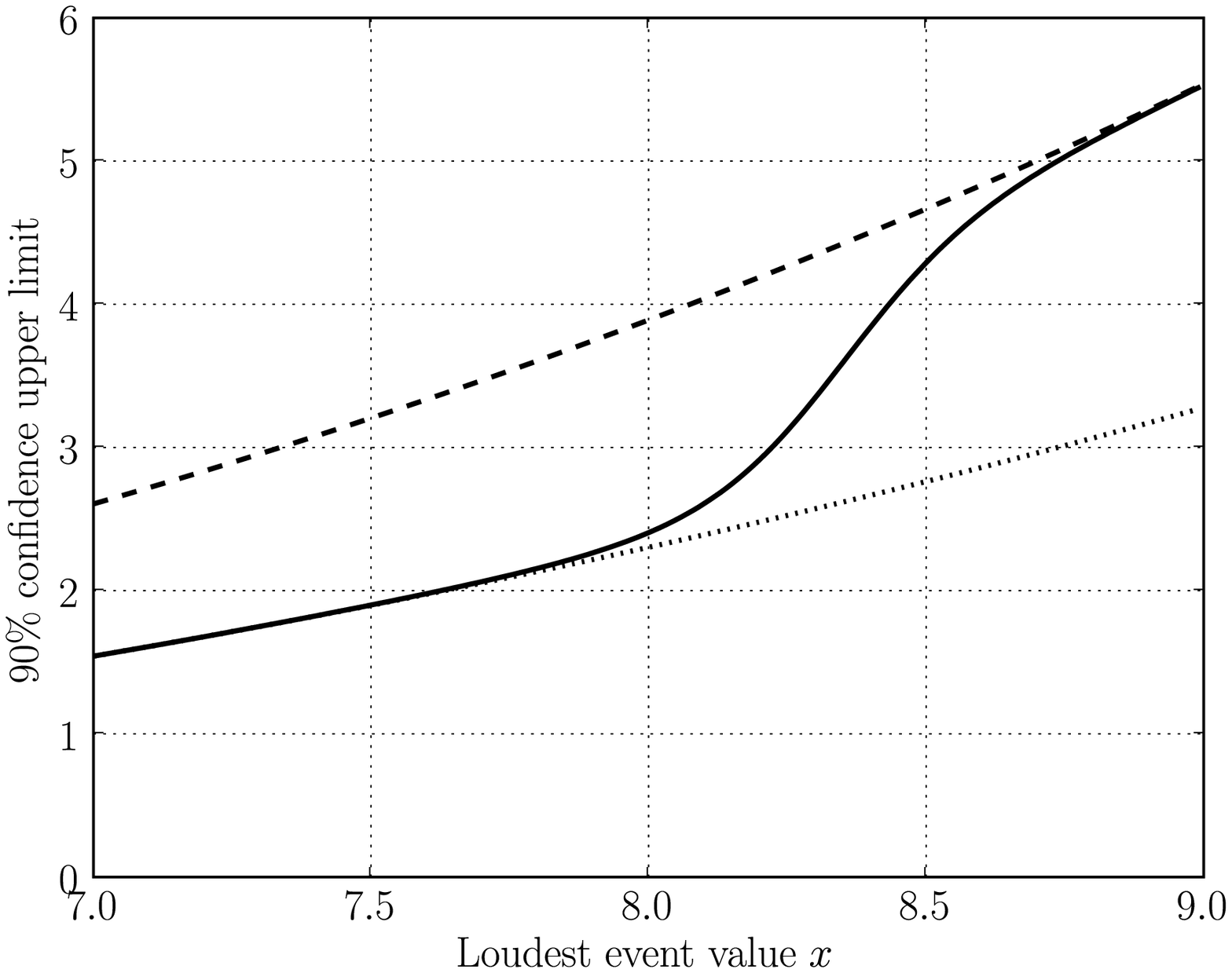}
\includegraphics[width=1.0\linewidth]{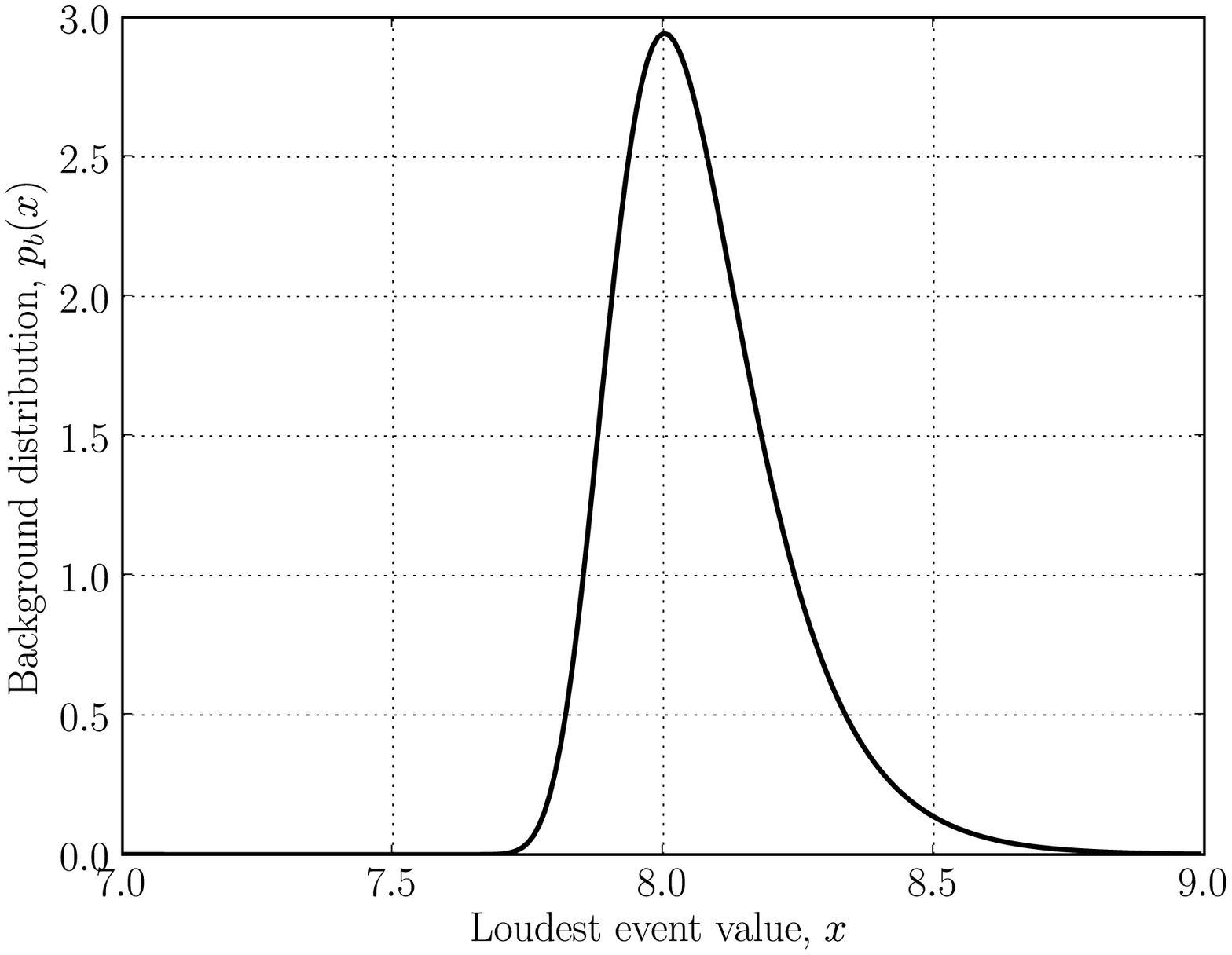}
\caption{\label{fig:loudest_example} 
a) The upper limit as a function of the observed loudest event.  The
solid line shows the value of the upper limit as a function of $\xm$.  The
dotted and dashed lines are given by $2.303/\epsilon(\xm)$ and $3.890/\epsilon(\xm)$.
We see that the upper limit transitions smoothly from one to the other.
At low values of $\xm$, the loudest event is very much consistent with the
background, $\Lambdam \approx 0$ and the upper limit is close to the
dotted line.  For larger values of $\xm$ the loudest event is more
consistent with foreground, $\Lambdam \rightarrow \infty$, and the upper limit
is more consistent with the dashed line.
b) The probability distribution for the loudest event assuming that it
is drawn from the background distribution, $p_0(x)$.  Multiplying the
upper limit curve by this distribution and integrating over $x$ gives
the expected value of the upper limit if the loudest event is from the
background.}  
\end{figure}

\begin{figure}[t]
\includegraphics[width=1.0\linewidth]{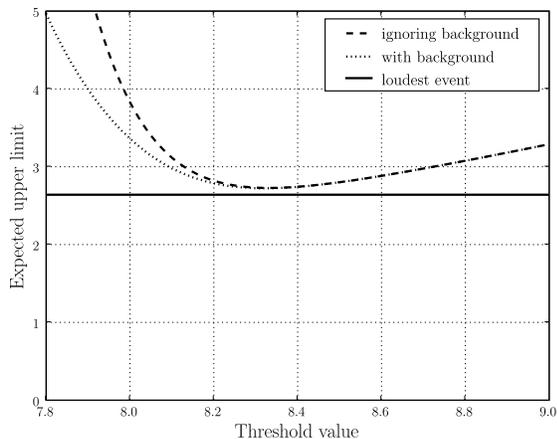}
\caption{\label{fig:loudest_vs_fixed} 
Figure showing the expected upper limit as a function of the fixed
threshold.  The dashed line shows the upper limit obtained when ignoring
the background, while the dotted line includes the background
contribution.  For large values of the threshold where the expected
background is small, both limits approach $2.303/\epsilon(\xt)$ as expected.
For low values of $\xt$, there is a good chance of many events above
threshold which leads to a worse upper limit.  The balance occurs at
around a threshold value of $\xt = 8.3$.  For reference, we also plot a
horizontal line showing the expected upper limit from the loudest event.
Interestingly, the loudest event will, on average, outperform the fixed
threshold for any value of the threshold.} 
\end{figure}

Let us compare the loudest event statistic against a fixed threshold
approach.  The loudest event prescription can be applied to any form
of background, provided the required quantities in
Eq.~(\ref{eq:Lambda}) can be measured or estimated.  In many
experiments, one might expect the background events above a statistic
value $x$ to be Poisson distributed, with mean $\nu_0(x)$ where
$\nu_0$ is a non-increasing function of $x$.  Then, it follows
directly that
\begin{eqnarray}\label{eq:poisson_bg} 
P_0(x) &=& e^{-\nu_0(x)} \nonumber \\ 
p_0(x) &=& \left|\frac{d\nu_0(x)}{dx}\right|
e^{-\nu_0(x)} \nonumber \\ 
p_0(x) / P_0(x) &=&
\left|\frac{d\nu_0(x)}{dx}\right| \; .  
\end{eqnarray} 
We work with an example where the mean $\nu_0(x) = e^{(8^2-x^2)/2}$ 
and the
foreground is distributed as $\epsilon(x) = (8/x)^{3}$.  These
choices are natural in the context of a search for gravitational waves
from coalescing binaries where the background is $\chi^2$ distributed
with two degrees of freedom, while the foreground is uniformly
distributed in volume, and the signal strength (and hence statistic
value $x$) are inversely proportional to the distance
\cite{systematics}. The normalizations of these functions are chosen
for simplicity so that $\nu_0(8) = \epsilon(8) = 1$.  The main feature
of these distributions, however, is simply that they are both
decreasing functions of $x$, and that the background decreases more
rapidly than the foreground.  The value of the upper
limit as a function of the actual loudest event
is shown in Fig.~\ref{fig:loudest_example}a.
The upper limit transitions smoothly from the zero foreground limit (at
low values of $x$) to zero background limit (at large values of $x$).
Figure~\ref{fig:loudest_example}b shows the distribution $p_0(x)$.
This corresponds to the expected distribution of for the loudest event
if it is due to the background.  Then, by multiplying the upper limit by
the expected distribution for the loudest event and integrating, we
obtain the expected upper limit.  In this example it is $2.64$.

For comparison, the upper limit for a fixed threshold is presented in
Figure \ref{fig:loudest_vs_fixed}.  When calculating the upper limit for
a fixed threshold, one simply counts the number of events $\nm$ above the
chosen threshold $\xt$ and obtains a limit
\begin{equation}
  \frac{\mu_{90\%}}{T} = \frac{ F(\nm) }{ \epsilon(\xt)T }
\end{equation}
where $F(n)$ is a known function for each integer $n$ (see, for example,
\cite{Yao:2006px} for more details). In particular, when zero events are
observed above the threshold, $F(0) = 2.303$.  When performing a fixed
threshold search, it is possible to take into account the expected
background and, much as for the loudest event, neglecting to do so will
lead to a conservative result.  In Fig.~\ref{fig:loudest_vs_fixed}, we
show the expected upper limit as a function of the threshold.     

Clearly, in this example, the loudest event statistic is preferable to a
fixed threshold, as it will provide a better expected upper limit value
than the fixed threshold for \textit{any} value of the threshold (with
or without the background).  We note that this result is specific to the
details of the example under consideration; the key feature is that the
background rate is a very steep function of $x$.  Indeed, in
\cite{loudestGWDAW03}, the same example was considered, but with an
expected background of unity at $\xm=4.5$ rather than $\xm=8$, leading to a
small range of values where the fixed threshold does beat the loudest
event.  However, as emphasized in that paper the attraction of the
loudest event is that it is unnecessary to fix a threshold ahead of
performing the search --- the search itself determines the threshold.

\bibliography{../bibtex/iulpapers}

\begin{thebibliography}{14}
\expandafter\ifx\csname natexlab\endcsname\relax\def\natexlab#1{#1}\fi
\expandafter\ifx\csname bibnamefont\endcsname\relax
  \def\bibnamefont#1{#1}\fi
\expandafter\ifx\csname bibfnamefont\endcsname\relax
  \def\bibfnamefont#1{#1}\fi
\expandafter\ifx\csname citenamefont\endcsname\relax
  \def\citenamefont#1{#1}\fi
\expandafter\ifx\csname url\endcsname\relax
  \def\url#1{\texttt{#1}}\fi
\expandafter\ifx\csname urlprefix\endcsname\relax\def\urlprefix{URL }\fi
\providecommand{\bibinfo}[2]{#2}
\providecommand{\eprint}[2][]{\url{#2}}

\bibitem[{\citenamefont{Allen et~al.}(1999)}]{Allen:1999yt}
\bibinfo{author}{\bibfnamefont{B.}~\bibnamefont{Allen}} \bibnamefont{et~al.},
  \bibinfo{journal}{Phys. Rev. Lett.} \textbf{\bibinfo{volume}{83}},
  \bibinfo{pages}{1498} (\bibinfo{year}{1999}).

\bibitem[{\citenamefont{Abbott et~al.}(2004{\natexlab{a}})}]{LIGOS1iul}
\bibinfo{author}{\bibfnamefont{B.}~\bibnamefont{Abbott}} \bibnamefont{et~al.}
  (\bibinfo{collaboration}{LIGO Scientific Collaboration}),
  \bibinfo{journal}{Phys. Rev. D} \textbf{\bibinfo{volume}{69}},
  \bibinfo{pages}{122001} (\bibinfo{year}{2004}{\natexlab{a}}),
  \eprint{gr-qc/0308069}.

\bibitem[{\citenamefont{Abbott et~al.}(2005{\natexlab{a}})}]{LIGOS2macho}
\bibinfo{author}{\bibfnamefont{B.}~\bibnamefont{Abbott}} \bibnamefont{et~al.}
  (\bibinfo{collaboration}{LIGO Scientific Collaboration}),
  \bibinfo{journal}{Phys. Rev. D} \textbf{\bibinfo{volume}{72}},
  \bibinfo{pages}{082002} (\bibinfo{year}{2005}{\natexlab{a}}),
  \eprint{gr-qc/0505042}.

\bibitem[{\citenamefont{Abbott et~al.}(2005{\natexlab{b}})}]{LIGOS2iul}
\bibinfo{author}{\bibfnamefont{B.}~\bibnamefont{Abbott}} \bibnamefont{et~al.}
  (\bibinfo{collaboration}{LIGO Scientific Collaboration}),
  \bibinfo{journal}{Phys. Rev. D} \textbf{\bibinfo{volume}{72}},
  \bibinfo{pages}{082001} (\bibinfo{year}{2005}{\natexlab{b}}),
  \eprint{gr-qc/0505041}.

\bibitem[{\citenamefont{Abbott et~al.}(2006)}]{LIGOS2bbh}
\bibinfo{author}{\bibfnamefont{B.}~\bibnamefont{Abbott}} \bibnamefont{et~al.}
  (\bibinfo{collaboration}{LIGO Scientific Collaboration}),
  \bibinfo{journal}{Phys. Rev. D} \textbf{\bibinfo{volume}{73}},
  \bibinfo{pages}{062001} (\bibinfo{year}{2006}), \eprint{gr-qc/0509129}.

\bibitem[{\citenamefont{Abbott et~al.}(2004{\natexlab{b}})}]{Abbott:2003yq}
\bibinfo{author}{\bibfnamefont{B.}~\bibnamefont{Abbott}} \bibnamefont{et~al.}
  (\bibinfo{collaboration}{LIGO Scientific}), \bibinfo{journal}{Phys. Rev.}
  \textbf{\bibinfo{volume}{D69}}, \bibinfo{pages}{082004}
  (\bibinfo{year}{2004}{\natexlab{b}}), \eprint{gr-qc/0308050}.

\bibitem[{\citenamefont{Siemens et~al.}(2006)}]{Siemens:2006vk}
\bibinfo{author}{\bibfnamefont{X.}~\bibnamefont{Siemens}} \bibnamefont{et~al.},
  \bibinfo{journal}{Phys. Rev.} \textbf{\bibinfo{volume}{D73}},
  \bibinfo{pages}{105001} (\bibinfo{year}{2006}), \eprint{gr-qc/0603115}.

\bibitem[{\citenamefont{Brady et~al.}(2004)\citenamefont{Brady, Creighton, and
  Wiseman}}]{loudestGWDAW03}
\bibinfo{author}{\bibfnamefont{P.~R.} \bibnamefont{Brady}},
  \bibinfo{author}{\bibfnamefont{J.~D.~E.} \bibnamefont{Creighton}},
  \bibnamefont{and} \bibinfo{author}{\bibfnamefont{A.~G.}
  \bibnamefont{Wiseman}}, \bibinfo{journal}{Class. Quant. Grav.}
  \textbf{\bibinfo{volume}{21}}, \bibinfo{pages}{S1775} (\bibinfo{year}{2004}).

\bibitem[{\citenamefont{Cousins}(1994)}]{Cousins:1993gs}
\bibinfo{author}{\bibfnamefont{R.}~\bibnamefont{Cousins}},
  \bibinfo{journal}{Nucl. Instrum. Meth.} \textbf{\bibinfo{volume}{A337}},
  \bibinfo{pages}{557} (\bibinfo{year}{1994}).

\bibitem[{\citenamefont{Yellin}(2002)}]{Yellin:2002xd}
\bibinfo{author}{\bibfnamefont{S.}~\bibnamefont{Yellin}},
  \bibinfo{journal}{Phys. Rev.} \textbf{\bibinfo{volume}{D66}},
  \bibinfo{pages}{032005} (\bibinfo{year}{2002}), \eprint{physics/0203002}.

\bibitem[{\citenamefont{Yao et~al.}(2006{\natexlab{a}})}]{Yao:2006}
\bibinfo{author}{\bibfnamefont{W.-M.} \bibnamefont{Yao}} \bibnamefont{et~al.},
  \bibinfo{journal}{J. Phys. G: Nucl. Part. Phys.}
  \textbf{\bibinfo{volume}{33}}, \bibinfo{pages}{1}
  (\bibinfo{year}{2006}{\natexlab{a}}).

\bibitem[{\citenamefont{Feldman and Cousins}(1998)}]{Feldman:1997qc}
\bibinfo{author}{\bibfnamefont{G.~J.} \bibnamefont{Feldman}} \bibnamefont{and}
  \bibinfo{author}{\bibfnamefont{R.~D.} \bibnamefont{Cousins}},
  \bibinfo{journal}{Phys. Rev. D} \textbf{\bibinfo{volume}{57}},
  \bibinfo{pages}{3873} (\bibinfo{year}{1998}), \eprint{physics/9711021}.

\bibitem[{\citenamefont{Brady and Fairhurst}(2007)}]{systematics}
\bibinfo{author}{\bibfnamefont{P.}~\bibnamefont{Brady}} \bibnamefont{and}
  \bibinfo{author}{\bibfnamefont{S.}~\bibnamefont{Fairhurst}},
  \bibinfo{journal}{gr-qc}  (\bibinfo{year}{2007}), \eprint{arXiv:0707.2410}.

\bibitem[{\citenamefont{Yao et~al.}(2006{\natexlab{b}})}]{Yao:2006px}
\bibinfo{author}{\bibfnamefont{W.~M.} \bibnamefont{Yao}} \bibnamefont{et~al.}
  (\bibinfo{collaboration}{Particle Data Group}), \bibinfo{journal}{J. Phys.}
  \textbf{\bibinfo{volume}{G33}}, \bibinfo{pages}{1}
  (\bibinfo{year}{2006}{\natexlab{b}}).

\end{thebibliography}

\end{document}